# Coupling of Electromagnetism and Gravitation in the Weak Field Approximation

M. Tajmar[*]

*Vienna University of Technology, 1040 Vienna, Austria*

C. J. de Matos[†]

*Coimbra University, 3000 Coimbra, Portugal*

**Abstract**

Using the weak field approximation, we can express the theory of general relativity in a Maxwell-type structure comparable to electromagnetism. We find that every electromagnetic field is coupled to a gravitoelectric and gravitomagnetic field. Acknowledging the fact that both fields originate from the same source, the particle, we can express the magnetic and electric field through their gravitational respective analogues using the proportionality coefficient ***k***. This coefficient depends on the ratio of mass and charge and the ratio between the electromagnetic and gravitic-gravitomagnetic permittivity and permeability respectively. Although the coefficient is very small, the fact that electromagnetic fields in material media can be used to generate gravitational and gravitomagnetic fields and vice versa is not commonly known. We find that the coupling coefficient can be increased by massive ion currents, electron and nuclear spin-alignment. Advances in material sciences, cryogenic technology and high frequency electromagnetic fields in material media may lead to applications of the derived relationships.

[*] Postdoctoral Associate, Present Address: European Space Research and Technology Centre ESA-ESTEC, P.O. Box 299, 2200 AG Noordwijk, The Netherlands. E-mail: tajmar@bigfoot.com
[†] Graduate Associate, Present Address: European Space Research and Technology Centre ESA-ESTEC, P.O. Box 299, 2200 AG Noordwijk, The Netherlands. E-mail: cdematos@estec.esa.nl

## 1. Introduction

Neglecting effects due to space curvature and special relativity, the theory of general relativity which describes gravitation can be linearized, unveiling a structure comparable to the Maxwell equations, which describe electromagnetism. This weak field approximation splits gravitation into components similar to the electric and magnetic field. Although masses attract each other and equal electric charges repel, the comparable structure between both the gravitational and the electromagnetic field gives rise to a coupling between them because both fields originate from the same point source, the particle. In the case of the gravitational field, the source is the mass of the particle, whereas in the case of the electromagnetic field, the source is the charge of the particle. Moving the particle will both create a magnetic and, using the analogy term, a gravitomagnetic field simultaneously. Hence, the induction of a gravitational field and (or) a gravitomagnetic field using electromagnetic fields applied to material media must be possible. This fact is not commonly known in nowadays physics.

The splitting of gravitation in an electric and magnetic type component originates back to as early as 1983 when Heaviside[1] investigated how energy is propagated in a gravitational field. His proposed gravitational Poynting vector contained the magnetic component of gravitation, which is hidden in the tensor equations of Einstein's general relativity theory, published more then 30 years afterwards. Forward[2] first expressed the linearized general relativity equations in a Maxwell-structure and proposed experiments to detect gravitomagnetic and non-Newtonian gravitational fields[3]. Further experiments were proposed by Braginski[4] indicating an increasing interest of detecting effects related to general relativity using laboratory equipment. However, the proposed experiments suggested effects too small for being detected. Recently, several approaches to produce larger gravitomagnetic fields using the unique properties of superconductors appeared[5,6,7] following the first approaches from DeWitt and Ross of modifying the London equation to include the gravitomagnetic field[8,9].

The main objective of this paper is to show that every electromagnetic field is coupled to a gravitoelectric-gravitomagnetic field and that induction between both fields is possible. The derived coupling is generally valid and does not require special properties like superconductivity. This enables a new approach to think about experiments which could modify the gravitational fields in terrestrial laboratories. Successful means for producing significant non-Newtonian fields promise many spin-off applications presently only possible in the microgravity environment of space[10].

## 2. Weak Field Approximation

In deriving an analogy between gravitation and electromagnetism, the following assumptions have been made:

1. all motions are much slower than the speed of light to neglect special relativity
2. the kinetic or potential energy of all bodies being considered is much smaller than their mass energy to neglect space curvature effects
3. the gravitational fields are always weak enough so that superposition is valid
4. the distance between objects is not so large that we have to take retardation into account

The procedure of linearizing Einstein's theory of general relativity is included in most textbooks[11,12]. We start with Einstein's field equation:

$$R_{ab} - \frac{1}{2} g_{ab} R = \frac{8pG}{c^4} T_{ab} \qquad (1)$$

Due to assumptions (1) and (3), the metric can be approximated by

$$g_{ab} \cong \boldsymbol{h}_{ab} + h_{ab} \qquad (2)$$

where the greek indices $a$, $b$ = 0, 1, 2, 3 and $\boldsymbol{h}_{ab}$ = (+1, -1, -1, -1) is the flat spacetime metric tensor, and $|h_{ab}| \ll 1$ is the perturbation to the flat metric. Using this form of the metric, we can approximate the Ricci tensor and the Ricci scalar by

$$R_{ab} \cong -\frac{1}{2}\left(\frac{1}{c^2}\frac{\partial^2}{\partial t^2} - \Delta\right) h_{ab} \qquad (3)$$

$$R \cong \boldsymbol{h}^{ab} R_{ab} = \frac{1}{2}\left(\frac{1}{c^2}\frac{\partial^2}{\partial t^2} - \Delta\right) h \qquad (4)$$

where in obtaining Equation (3) and (4) we choose our coordinate system so that we have the following "gauge" condition

$$\left[h_{ab} - \frac{1}{2}\mathbf{h}_{ab}h\right]_{,b} = 0 \tag{5}$$

Defining the gravitational potential as

$$\bar{h}_{ab} = h_{ab} - \frac{1}{2}\mathbf{h}_{ab}h \tag{6}$$

we can substitute Equation (3) and (4) into (1) and get

$$\frac{1}{c^2}\frac{\partial^2}{\partial t^2}\bar{h}_{ab} - \Delta\bar{h}_{ab} = -\frac{16pG}{c^4}T_{ab} \tag{7}$$

## 3. The Einstein-Maxwell-Type Gravitational Equations

In this section we will derive the gravitational analogues to the scalar and vector potential used to express the Einstein-Maxwell equations. We will cover only the necessary results, for a detailed analysis the reader is referred to the literature[13,14].

In the first approximation (zero order for the energy momentum tensor), we assume that all quantities are not varying with time. Then the time derivative of the gravitational potential is zero and all components of the energy-momentum tensor are zero except

$$T_{00} = \mathbf{r}_m c^2 \tag{8}$$

where $\mathbf{r}_m$ is the mass density. Hence, the weak field approximation in Equation (7) reduces to the Poisson equation with the solution

$$\Phi_g = \frac{c^2 \bar{h}_{00}}{4} = -\frac{1}{4p\mathbf{e}_g}\int_V \frac{\mathbf{r}_m}{r}dV \tag{9}$$

where $F_g$ is the gravitational analog to the scalar potential and $e_g$ the gravitational permittivity which we define using Newton's gravitational constant $G$ as

$$e_g = \frac{1}{4\pi G} = 1.19 \times 10^9 \frac{\text{kg} \cdot \text{s}^2}{\text{m}^3} \tag{10}$$

In the next higher approximation (first order for the energy-momentum tensor), we still assume that the potential is not varying in time, but that the masses involved are moving at a certain velocity or rotate. Then the energy-momentum tensor will have the additional components

$$T_{0i} = -\rho_m c^2 \left(\frac{v_i}{c}\right) \tag{11}$$

Using again the weak field Equation (7) we can define the analog to the vector potential using the mass density flow $\vec{p} = \rho_m \vec{v}$ as

$$\vec{A}_g = -\frac{c}{4}\bar{h}_{0i} = -\frac{\mu_g}{4\pi} \int_V \frac{\vec{p}}{r} dV \tag{12}$$

where $\mu_g$ is the gravitomagnetic permeability defined by

$$\mu_g = \frac{4\pi G}{c^2} = 9.31 \times 10^{-27} \frac{\text{m}}{\text{kg}} \tag{13}$$

Here we use a different definition of the gravitomagnetic permeability than various other publications[2] so that

$$c = \frac{1}{\sqrt{e_g \mu_g}} \tag{14}$$

which expresses, that a gravitational wave travels at the speed of light in analogy to an electromagnetic wave[15]. Defining the gravitoelectric field $\vec{g}$ and the gravitomagnetic field $\vec{B}_g$ by

$$\vec{g} = -\nabla \Phi_g - \frac{\partial \vec{A}_g}{\partial t} \quad \left[\frac{m}{s^2}\right] \qquad (15)$$

$$\vec{B}_g = rot\, \vec{A}_g \quad \left[\frac{1}{s}\right]$$

we can rearrange the weak field Equation (7) to reassemble a Maxwellian structure called the Maxwell-Einstein equations. Our initial assumption of a non time-varying gravitational potential $\bar{h}_{ab}$ is not necessary in this rearrangement allowing a time dependent $\vec{A}_g$, however, the identification of the gravitational scalar and vector potential is much more complicated. For a detailed derivation the reader is referred to the cited references at the beginning. We now compare the obtained Maxwell-Einstein equations to the standard electromagnetic Maxwell equations:

$$\begin{array}{ll}
div\, \vec{E} = \dfrac{r}{e_0} & div\, \vec{g} = -\dfrac{r_m}{e_g} \\
div\, \vec{B} = 0 & div\, \vec{B}_g = 0 \\
rot\, \vec{E} = -\dfrac{\partial \vec{B}}{\partial t} & rot\, \vec{g} = -\dfrac{\partial \vec{B}_g}{\partial t} \\
rot\, \vec{B} = m_0\, r\vec{v} + \dfrac{1}{c^2}\dfrac{\partial \vec{E}}{\partial t} & rot\, \vec{B}_g = -m_g\, r_m\vec{v} + \dfrac{1}{c^2}\dfrac{\partial \vec{g}}{\partial t}
\end{array} \qquad (16)$$

Maxwell Equations      Maxwell-Einstein Equations
(Electromagnetism)      (Gravitation)

The volume integrals in Equation (9) and (12) assume point masses. These integrals would have to be modified in the case of masses with considerable volumes. However, in the following we will only concentrate on the ratio between gravitational and electromagnetic scalar and vector potentials. Hence, the integral will cut and has no further importance in our outlined coupling concept.

## 4. Coupling between Electromagnetism and Gravitation

Equation (16) shows the very similar structure between the electromagnetic and the linearized gravitational field. We neglected the effects of spacetime curvature from heavy masses which is a direct result from the strong equivalence principle and special relativistic effects. Those effects would be prominent in the case of strong gravitational fields or velocities near the speed of light, however, the structure given by Equation (16) would still apply even with a minor contribution in extreme cases.

Considering our definition of the gravitational scalar and vector potential in Equation (9) and (12), we can propose the following relationship valid for equal charged particles:

$$\boldsymbol{r}_m = \frac{m}{e}\boldsymbol{r} \tag{17}$$

$$\Phi_g = -\frac{m}{e}\frac{e_0}{e_g}\Phi \tag{18}$$

$$\vec{A}_g = -\frac{m}{e}\frac{m_g}{m_0}\vec{A} \tag{19}$$

where $e$ is the charge and $m$ is the mass of the particle if we assume the same volume of integration for Equation (12) and its magnetic analogue. Considering a moving ion, the $m/e$ ratio can be several orders of magnitude higher than for a single electron due to the much higher mass. If we look at the coefficients between the gravitic-gravitomagnetic and electromagnetic scalar and vector potentials, we see that they are equal:

$$\frac{e_0}{e_g}\frac{m_0}{m_g} = 1 \Leftrightarrow \frac{e_0}{e_g} = \frac{m_g}{m_0} \tag{20}$$

By defining the proportionality coefficient $\boldsymbol{k}$

$$\boldsymbol{k} = -\frac{m}{e}\frac{\boldsymbol{m}_g}{\boldsymbol{m}_0} = -\frac{m}{e}\frac{\boldsymbol{e}_0}{\boldsymbol{e}_g} = -7.41x10^{-21}\cdot\frac{m}{e} \tag{21}$$

and putting equations (18), (19) and (21) into equations (15), we can now express the electromagnetic components by gravitomagnetic ones and vice versa:

$$\begin{aligned} \boldsymbol{f}_g &= \boldsymbol{k} \cdot \boldsymbol{f} \\ \vec{A}_g &= \boldsymbol{k} \cdot \vec{A} \\ \vec{g} &= \boldsymbol{k} \cdot \vec{E} \\ \vec{B}_g &= \boldsymbol{k} \cdot \vec{B} \end{aligned} \qquad (22)$$

Both the electromagnetic and the gravitational field originate from the same source, the particle. Hence, moving the source will create both electromagnetic and at the same time gravitoelectric-gravitomagnetic fields. Considering only particles with equal charges, the field lines of the magnetic and gravitomagnetic field ($\vec{B}$ and $\vec{B}_g$) and of the electric and gravitoelectric field ($\vec{E}$ and $\vec{g}$) would look completely similar being proportional or indirect proportional to the coefficient $\boldsymbol{k}$ given by equation (21).

However, the coupling is very small. For the case of an electron moving in a vacuum environment, the proportionality coefficient $\boldsymbol{k}=4.22 \times 10^{-32}$ $T^{-1}s^{-1}$. Hence, the magnetic field produced by the movement of equal charges is associated with a very weak gravitomagnetic field being more than 32 orders of magnitude below. This coefficient can be increased if we think about a higher $m/e$ ratio, for example using protons ($\boldsymbol{k}_{proton} = -7.6 \times 10^{-29}$ $T^{-1}s^{-1}$), ions (e.g. single charged positive lead ions $\boldsymbol{k}_{lead} = -1.6 \times 10^{-26}$ $T^{-1}s^{-1}$) or electrically charged clusters with even higher masses.

We find that every electromagnetic field is associated with a gravitic-gravitomagnetic field. However, due to the possibility of having neutral matter, only the gravitomagnetic field can exist without an associated electromagnetic field.

## 5. Non-Newtonian Gravity using Electromagnetic Fields

The divergence part of Equation (16) describes the well known fact, that a mass density (or an electric charge density associated with its mass) produces a gravitational field. The remaining rotational part, however, unveils an additional way of creating gravitational fields similar to Faraday's electromagnetic induction principle (non-Newtonian gravity).

Using the relationship between the magnetic and gravitomagnetic field in Equation (22), we can rewrite this part of the Einstein-Maxell-type equations leading to

$$rot\,\vec{E} = -\frac{1}{k}\frac{\partial \vec{B}_g}{\partial t}$$

$$rot\,\vec{B} = \frac{e}{m}\boldsymbol{m}_0\,\boldsymbol{r}_m\vec{v} + \frac{1}{k}\frac{1}{c^2}\frac{\partial \vec{g}}{\partial t} \qquad (23)$$

Coupled Maxwell-Einstein Equations

(Gravitation→Electromagnetism)

$$rot\,\vec{g} = -k\frac{\partial \vec{B}}{\partial t}$$

$$rot\,\vec{B}_g = -\frac{m}{e}\boldsymbol{m}_g\,\boldsymbol{r}\vec{v} - k\frac{1}{c^2}\frac{\partial \vec{E}}{\partial t} \qquad (24)$$

Coupled Maxwell-Einstein Equations

(Electromagnetism→Gravitation)

Equation (23) can be used to detect gravitational waves (assuming single charged matter) whereas Equation (24) indicates an induction of non-Newtonian gravitational fields using electromagnetic fields. Due to the small value of **k**, present day technology capable of constantly producing magnetic fields in the order of 10 Tesla using large coils with high electron currents at a frequency of 1 GHz would only create a maximum gravitoelectric fields of $\vec{g}$ =2.65x10$^{-21}$ ms$^{-2}$. Hence, it looks very discouraging using the obtained relationships to design experiments which create measurable non-Newtonian gravitational fields.

Looking at the coefficient **k**, we suggest two possible ways of increasing the coupling which may result in stronger non-Newtonian gravitational fields:

1. <u>increasing the *m/e* ratio:</u> As already outlined, massive ions instead of electrons can increase this ratio by 6 orders of magnitude (e.g. single charged lead ions). This could be achieved using dense plasmas which are accelerated by electrostatic/magnetic fields to gain high velocities.

2. <u>gravitomagnetic analogue to magnetism:</u> Similar to para-, dia-, and ferro-magnetism, the angular and spin momentums from free electrons in material media could be used to propose a gravitomagnetic relative permeability $m_{gr}$ which could increase the gravitomagnetic field $\vec{B}_g$. If we define the gravitomagnetic electron spin momentum $\vec{m}_g$ and the corresponding gravitomagnetic magnetization $\vec{M}_g$ as:

$$\vec{m}_g = -\frac{m}{e}\vec{m} \quad \left[\frac{kgm^2}{s}\right]$$

$$\vec{M}_g = \frac{\vec{m}_g}{V} = -\vec{M}\cdot\frac{m}{e} \quad \left[\frac{kg}{ms}\right]$$
(25)

where $\vec{m}$ is the electron spin momentum and $\vec{M}$ is the magnetization, we can express the gravitomagnetic susceptibility $c_g$ using

$$\vec{M}_g = c_g \cdot \frac{\vec{B}_g}{m_g}$$
(26)

If we substitute equation (26) with (25) and (22), we get

$$c_g = c$$
(27)

We see, that the gravitomagnetic field will be increased by the momentum alignment in materials through the presence of an external magnetic field with the same susceptibility than in the case of magnetism. On the other hand, the gravitomagnetic susceptibility of matter if only exposed to gravitomagnetic fields is neglectible.

*Para- and Diamagnetism*

If we align the spin momentums only with gravitomagnetic fields, we can't use the correlation between the magnetic and gravitomagnetic momentums given by Equation (25). In this case, we have to look at the expression for the magnetic susceptibility using spin moments as described in solid-state physics textbooks[16]. We then exchange the electromagnetic terms using the analog gravitic-gravitomagnetic terms to express the gravitomagnetic susceptibility. We find that for para- and diamagnetism, the magnetic susceptibility scales as

$$c \propto m_0 \cdot \vec{m}^2 \tag{28}$$

We can then estimate the value of $c_g$ using Equation (25)

$$\frac{c_g}{c} = \frac{m_g}{m_0} \cdot \frac{m^2}{e^2} \tag{28}$$

Considering electron spins, the gravitomagnetic susceptibility relates to the magnetic susceptibility as $c_g / c = 2.39 \times 10^{-43}$.

*Ferromagnetism*

The temperature associated with spontaneous magnetic spin ordering is proportional to the square of the magnetic (in this case gravitomagnetic) momentum[16]. Therefore applying the previous result for gravitomagnetic spin ordering, we expect temperatures in the range of 300 K*$m^2/e^2$ which is $9.7 \times 10^{-21}$ K for an electron, extrapolating from normal ferromagnetic phenomena at room temperatures.

*Nuclear Magnetism*

Another contribution to gravitomagnetism can result from nuclear momentums. In electromagnetism, nuclear magnetism is very small compared to para-, dia-, and ferro-magnetism due to the higher nuclear mass compared to electron's mass. However, nuclear gravitomagnetic momentums must have the same order of magnitude than gravitomagnetic electron spins considering its definition in Equation (25). Therefore, nuclear magnetism will influence gravitomagnetism at the same order of magnitude than normal magnetism at very low temperatures required to align the nuclear spins (typically at $10^{-9} - 10^{-3}$ K) by an applied magnetic field.

Due to the absence of negative masses, there is no gravitomagnetic analogue to the polarization $\vec{P}$ which defines the relative permittivity $e_r$. Hence, the relative gravitational permittivity in material media is always equal to 1. Therefore, the construction of a gravitational "Faraday-Cage" to shield gravitational fields is impossible.

## 6. Conclusion

The presented coupling of electromagnetism and gravitation in the weak field approximation shows that both fields can be converted into each other. We find that every electromagnetic field is associated with a gravitomagnetic field, only gravitomagnetic fields can exist without electromagnetic ones in the case of neutral matter. Due to the small coupling, experiments which modify the gravitational field in a terrestrial laboratory will involve a lot of engineering to achieve measurable results.

Comparing the non-Newtonian gravitational acceleration to the one of a 1 kg point mass at a radius of 10 cm ($\vec{g} = 6.7 \times 10^{-9}$ ms$^{-2}$), we see that the fields associated with electromagnetism are usually at least 10 orders of magnitude below the gravitational fields produced by standard masses. However, electromagnetism is associated with available wide ranges of frequencies, currents and potentials and is therefore more "engineerable" than the handling of heavy masses. Hence, using electromagnetic fields to induce non-Newtonian gravitational fields is an issue closely related to material sciences, cryogenic technology and extremely high frequency problems.

Additionally to this, we find that massive ion currents, electron and nuclear spin-alignment in materials can lead to a higher coupling coefficient. The presented relationship encourages further research into practical applications towards experimental gravitation.

Our proposed coupling concept between gravitation and electromagnetism shall help to understand the similarity of both force fields despite their different mathematical description in general relativity and Maxwell theory. This is necessary along the goal of physics to unify all forces in nature.

**References**

[1]Heaviside, O., "A Gravitational and Electromagnetic Analogy," The Electrician, **31**, 281-282 and 359 (1893)

[2]Forward, R.L. "General Relativity for the Experimentalist," Proceedings of the IRE, **49**, 892-586 (1961)

[3]Forward, R.L., "Guidelines to Antigravity," American Journal of Physics, **31**, 166-170 (1963)

[4]Braginski, V.B., Caves, C.M., Thorne, K.S., "Laboratory Experiments to Test Relativity Gravity," Physical Review D, **15**(5), 2047-2068 (1977)

[5]Li, N., and Torr, D.G., "Effects of a Gravitomagnetic Field on Pure Superconductors," Physical Review D, **43**(2), 457-459 (1991)

[6]Li, N., and Torr, D.G., "Gravitational Effects on the Magnetic Attenuation of Superconductors," Physical Review B, **46**, 5489 (1992)

[7]Li, N., Noever, D., Robertson, T., Koczor, R., Brantley, W., "Static Test for a Gravitational Force Coupled to Type II YBCO Superconductors," Physica C, **281**, 260-267 (1997)

[8]DeWitt, B.S., "Superconductors and Gravitational Drag," Physical Review Letters, **16**, 1092 (1966)

[9]Ross, D.K., "The London Equation for Superconductors in a Gravitational Field," *J. Phys. A: Math Gen*, **16**, 1331-1335 (1983)

[10]Walter, H.U., Fluid Sciences and Materials Science in Space: A European Perspective, (Springer Verlag, 1987)

[11]Møller, C., The Theory of Relativity, (Oxford University Press, London, 1952)

[12]Misner, C.W., Thorne, K.S., and Wheeler, J.A., Gravitation, (W.H.Freeman, San Francisco, 1973)

[13]Campbell, W.B., Morgan, T.A., "Maxwell Form of the Linear Theory of Gravitation," American Journal of Physics, **44**, 356-365 (1976)

[14]Peng, H., "On Calculation of Magnetic-Type Gravitation and Experiments," General Relativity and Gravitation, **15**(8), 725-735 (1983)

[15]Jefimenko, O.D., Causality, Electromagnetic Induction and Gravitation, (Electret Scientific Company, 1992)

[16]Ali Omare, M., Elementary Solid-State Physics, (Addison-Wesley Publication Company, 1996)